\documentclass[preprint2]{aastex} 

\usepackage{natbib}
\usepackage{graphicx}
\usepackage{bm}

\def\Mpc{\,h^{-1}\,{\rm Mpc}}

\def\s8{$\sigma_8$}

\def\eff{{\rm eff}}

\begin{document}

\title{Weakly nonlinear dynamics and the \s8 parameter}

\author{Roman Juszkiewicz$^{a,b}$, Hume A. Feldman$^{c,d,e}$, J. N. Fry$^f$, Andrew H. Jaffe$^e$}
\affil{$^a$Institute of Astronomy, 65-516 Zielona G\'{o}ra}
\affil{$^b$Copernicus Astronomical Center, 00-716 Warsaw, Poland}
\affil{$^c$Department of Physics and Astronomy, University of Kansas, Lawrence KS 66045, USA}
\affil{$^d$Department of Physics and Astronomy, University College London, London, WC1E 6BT, UK}
\affil{$^e$Astrophysics, Blackett Laboratory, Imperial College, London SW7 2AZ, UK}            
\affil{$^f$Department of Physics, University of Florida, Gainesville FL 32611-8440, USA}
\email{Email: roman@camk.edu.pl, feldman@ku.edu, fry@phys.ufl.edu, a.jaffe@imperial.ac.uk}

\begin{abstract}

The amplitude of cosmological density fluctuations, \s8, has been
studied and estimated by analysing many cosmological 
observations. The values of the estimates vary considerably between 
the various probes. However, different estimators probe the value of
\s8 in different cosmological scales and do not take into account the nonlinear
evolution of the parameter at late times. We show that estimates of the
amplitude of cosmological density fluctuations derived from cosmic flows are systematically
higher than those inferred at early epochs from the CMB because of nonlinear evolution
at later times. We discuss the past and future evolution of linear and nonlinear
perturbations, derive corrections to the value of \s8  and compare
amplitudes after accounting for these differences.
\end{abstract}

\keywords{cosmic flows, cosmological parameters from LSS, cosmological parameters from CMBR, galaxy dynamics}

\section{Introduction} 
\label{sec:introduction}

The $\sigma_8$ parameter is one of the most important and least well
known parameters in cosmology. It is a convenient measure of the
degree of the inhomogeneity of the Universe. 
It is the rms matter density contrast in a sphere with a comoving
radius of  $8 h^{-1}\;\textrm{Mpc}$ at present,
where $h$ is the usual dimensionless Hubble constant in
units of 100 km s$^{-1}$ Mpc$^{-1}$.
The original motivation for the use of this parameter was the need to
define the clustering amplitude: 
the variance in the number density of optical galaxies in $8h^{-1}\;\textrm{Mpc}$ spheres, $\sigma^2_{\rm gal}$,
was observed to be unity  \citep{DavPee83,StrWil95}. This parameter is
simply related to scale-independent linear bias 
\citep{Kai88}, defined as the square root of the ratio of the number
density variance to the 
mass variance  $b\equiv\sigma_{\rm gal}/\sigma_{\rm mass}=1/\sigma_8$.
Later, \s8 provided the standard way to normalize the theoretical power spectra of density fluctuations, determined
from first principles only up to an arbitrary multiplicative constant.
A theory can be tested against the empirical 
data by comparing its assumed value of $\sigma_8$ with a value implied by observations.
A comparison of values of $\sigma_8$, estimated from
various observations can provide an 
important measure of systematic errors introduced by different estimators,  
based on different physical effects, e.g., bulk flows, cosmic
microwave background (CMB) 
fluctuations or gravitational lensing. Such a comparison also provides
an internal consistency test for the gravitational instability theory.

To compare the $\sigma_8(z)$, estimated at high redshift, $z$, with its current value, estimated from observations at $z = 0$, 
it is necessary to take the growth of the fluctuations into account. Since all known estimates
of this parameter were of the order of unity or less, until recently the linear perturbation theory was used for this purpose:
\begin{equation}
\sigma_8(0) \, \approx \, \sigma_8 (z) \, {D(0) \over D(z)} \; \; ,
  \label{eq:linear}
\end{equation}
where $D$ is the linear growth factor \citep{pjep}.  
As we show below, ignoring higher order terms in the above expression introduces a systematic
error of order of 10 to 15\%. With the latest improvements in the quality of observations, such efects should not be neglected
and we provide a simple recipe how to take them into account. We focus on the comparison
of $\sigma_8$ estimated from the peculiar velocity field at an effective redshift $z=0$, to \s8 implied by the
CMB (effective redshift $z \approx 1100$). 

Relative motions of pairs of galaxies \citep{pairwise} as well as bulk
flows \citep{pikhud05,sarfelwat07} and velocity shear \citep{watfel07,felwat08} measurements have been
used to estimate $\sigma_8$ and $\Omega_m$, the density of the nonrelativistic matter. 
In particular, from pairwise velocities we found in \citet{pairwise} 
\begin{equation}
\sigma_8 = 1.13^{+0.22}_{-0.23} \; .
  \label{eq:v12}
\end{equation}
From CMB temperature fluctuations, the WMAP collaboration \citep{WMAP5} found 
\begin{equation}
\sigma_8 =  0.80 \pm 0.04 \; ,
  \label{eq:wmap}
\end{equation}
described  as the ``linear theory amplitude''  (see their Table 1). 
The above two estimates differ only slightly, at the level of
1.5-$\sigma$, and it is a success of the model that inferences 
by such different methods applied at two greatly different epochs are in good agreement.

In this paper we bring the two estimates even closer by  taking 
nonlinear dynamics into account. We find that the nonlinear correction
is modestly significant, given the improving accuracy in cosmology:
the linear value of $\sigma_8$, hereafter 
denoted $\sigma_L$, can be smaller by $\sim10$\% than the nonlinear
value, a systematic difference that is comparable to 
the current statistical uncertainties. In section~\ref{sec:window} we
describe the spatial window functions used 
to define \s8. In  section \ref{sec:linear_and_nonlinear_amplitudes}
we provide the recipe for recovering $\sigma_L$ from $\sigma$. 
In section~\ref{sec:discussion} we compare our revised \s8
parameters to other observational estimates of \s8. We summarize our results in section~\ref{sec:conc}.

\section{Window functions} 
\label{sec:window}

Like many cosmological experiments, measurements of cosmic flows are
sensitive to a windowed integral of the matter 
power spectrum. In general, such an observable can be characterized as
\begin{equation}
  Q_i = \int_0^\infty dk\; W(k) P_p(k) T_i^2(k)
  \label{eq:windowedPk}
\end{equation}
where $P_p(k)$ gives the primordial power spectrum, 
$k$ is the comoving wavenumber, $T_i^2(k)$ gives the transfer function
which contains the physics of the evolution 
of the particular observable from the primordial spectrum. The window
function, $W(k)$, describes the experimental 
setup (sky coverage, depth, errors etc.). This formalism describes
straightforward measurements 
of the galaxy power spectrum, in which case $Q_i=P_{\rm gal} (k_i)$,
the CMB spectrum for which $Q_i = C_{\ell_i}$, and the 
amplitude of the cosmological velocity field where $Q_i=P_{v} (k_i)$,
the velocity power spectrum. It is crucial 
to note that the transfer function depends implicitly upon the other
cosmological parameters 
and hence any lack of knowledge thereof will (or at least should) translate to increased uncertainty upon the amplitude.

Bulk flow and shear measure the velocity-velocity power spectrum [or
covariance, \cite{WatFelHud09,FelWatHud09}], 
whereas pairwise velocities measure the density-velocity
cross-spectrum 
\citep{pairwise}. Under \emph{linear} evolution, the density contrast
is proportional to the divergence of the peculiar velocity in real
space, or $\mathbf{v}\propto\mathbf{k}\rho$ in 
Fourier space, so these power spectra differ by powers of wavenumber
$k$ from the density power spectrum, which can be absorbed into the appropriate transfer function.

An amplitude parameter such as $\sigma_8$ is in essence a spectral
observable as well. We define the \emph{window function}
$W_R(\mathbf{x},  \mathbf{x'})=W_R(\mathbf{x-x'})$, 
normalized so that $\int d^3x \; W_R(x) = 1$. For our spherical top
hat, $W_R(r) = 1/V$ where $V = 4\pi R^3/3$, 
when $r=|\mathbf{x'-x}|\le R$, 0 otherwise.  Hence, the density
contrast, spatially averaged 
over a sphere around a particular point $\mathbf{x}$ is simply 
\begin{equation}
  \delta_R(\mathbf{x}) = \int d^3 x' W_R(\mathbf{x-x'})\delta(\mathbf{x'})\; \; .
\end{equation}
The ensemble avarage of $\delta_R^2$ at redshift $z$ is given by
\begin{equation}
  \sigma^2_R (z) \equiv \langle \delta_R^2 \rangle = \int_0^\infty \frac{dk}{k} \,\Delta(k,z)^2|{\tilde W}(kR)|^2\; ,
\end{equation}
where 
\begin{equation}
{\tilde W}(kR) = (3/kR)\,j_1(kR)\;  
\end{equation}
is the Fourier transform of the window function $W_R(r)$, and $j_1$ is a spherical Bessel
function of the first kind, while
\begin{equation}
\Delta \equiv 4\pi P(k,z)k^3/(2\pi)^3  
\end{equation}
is the dimensionless power per wavenumber octave and $P(k,z)$ is the power spectrum of the mass density fluctuations.
The mean square fluctuation that we would measure from the actual
density contrast depends upon the actual nonlinear 
power spectrum, but we can analogously define the linear variance by
replacing the power spectrum in the above expression with the 
linear spectrum, constrained to have the same amplitude for $k\to0$ (i.e., on large scales where nonlinear evolution is negligible).

The spatial representation of the above integral is given by the expression
\begin{equation}
\sigma^2_R(z) = \frac1{V^2} \int_V d^3 x\,d^3x' \, 
	\xi(|\mathbf{x} - \mathbf{x'}|, z) \,,  \label{eqn:sigmaDef}
\end{equation}
where $\xi$ is the two-point correlation function, and $V = 4\pi
R^3/3$. In \citet{pairwise} we have used the above equation to
estimate the 
true present-day value of the $\sigma_8$ parameter from the PSC$z$
survey correlation function \citep{hamteg02}. We have also 
used other empirical correlation functions, derived from different
surveys, as a template and we found that the resulting value of 
\s8 was not sensitive to such variations; see \citet{pairwise}.

\section{Linear and nonlinear amplitudes}
\label{sec:linear_and_nonlinear_amplitudes}

We use two different methods to estimate the nonlinear 
corrections for $\sigma_8$, one based in perturbation theory, which
allows us to express the correction as a simple analytical expression, 
and one using a phenomenological mapping based on conservation of pair
counts and calibrated using numerical simulations, 
which allows us to explore the effects of changing many parameters individually. 

\subsection{Linear perturbation theory}
\label{sec:PT}

\begin{figure}
			\includegraphics[width=1\columnwidth]{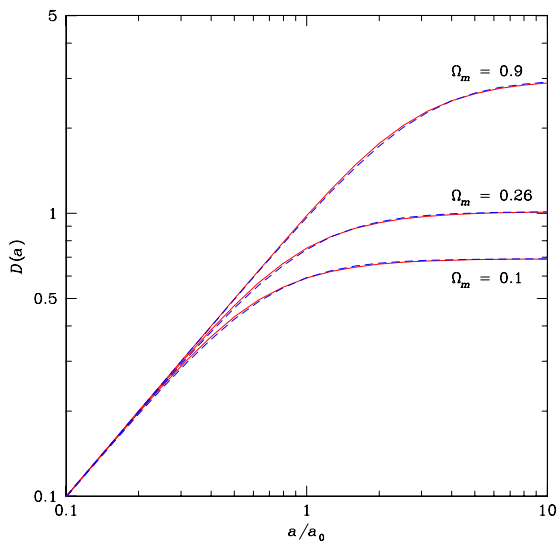}
\caption{\label{fig:Dofa}
The growth factor $D$, plotted as a function of the scale factor $a$ for three different values of $\Omega_m$. 
The exact solution~(\ref{eqn:Dexact}) is represented by the solid red
curve, while the dashed blue curve was derived from the approximate
formula~(\ref{eqn:Dapprox}). All models assume zero spatial curvature,
$\Omega_{\Lambda} = 1 - \Omega_m$. The scale factor at present equals
unity. Hence the past corresponds to $a/a_0 < 1$, while the future to $a/a_0 > 1$ ($a_0\equiv a(z=0)=1$).
}
\end{figure}

Under linear evolution, the spatial and temporal dependence of clustering separates. The density perturbation $\delta({\bf x}, a)=\delta\rho/\rho$ 
can be described as \citep{pjep}
\begin{equation}
\delta({\bf x},a) = \delta^{(1)} ({\bf x)}D(a)\;\;,
\label{eqn:L}
\end{equation}
where $\delta^{(1)}$ gives the linear density perturbation 
field as a function of comoving spatial coordinates ${\bf x}$ at some fiducial time and $D$ is the growth function, here parameterized by
the scale factor $a$ as a time coordinate (here and below we keep only the fastest-growing modes). In flat $\Lambda$CDM models 
the growing mode is given by \citep{Heath:1977p777, pjep2}
\begin{equation}
D(a) = \frac{5\Omega_m E(a) }{ 2} \, \int_0^a \, \frac{ du}{ u^3 E^3 (u)} \;\; ,
\label{eqn:Dexact}
\end{equation}
where
\begin{equation}
E(a) \equiv \left[ \, \Omega_m a^{-3} + 1 - \Omega_m \,\right]^{1/2} \; ,
\label{eqn:EofZ}
\end{equation}
In  the early Universe, when the scale factor is small, $a \to 0$,
equation~(\ref{eqn:Dexact}) is well approximated by the expression
\begin{equation}
D(a) = a \;\;,
\label{eqn:EdeSlimit}
\end{equation}
as in Einstein-de Sitter Universe. In the opposite limit, the cosmological constant
becomes dynamically dominant  and the linear growth factor will saturate at a maximum value, 
as gravitational clustering is balanced by the effective force of accelerated expansion. 
It is easy to show that in the limit $a\to\infty$,  the growth factor
is given by the expression
\begin{eqnarray}
D_{\infty} 	 =
        \frac{2\Gamma({2\over3})\Gamma({11\over6})}{\sqrt{\pi}}
        \left(\frac{\Omega_m}{1 - \Omega_m}\right)^{1/3} \;\; .
\label{eqn:Dinf}
\end{eqnarray}
Using the above two asymptotic expressions we have found 
a new fitting formula for the growth factor, valid for all flat $\Lambda$CDM cosmological models:
\begin{equation}
	D(a) = \frac{a}{\left[1+\left(a/D_{\infty}\right)^{2.3}\right]^{1/2.3}}\;\;.
\label{eqn:Dapprox}
\end{equation}
In Figure~\ref{fig:Dofa} we show that  equation~(\ref{eqn:Dapprox}) remains within two-percent level
of the exact solution~(\ref{eqn:Dexact}) in both the past and the future.
Note that some expressions for $D(a)$ and its logarithmic
derivative, $d \log D/d\log a$,  frequently quoted in the literature 
\citep{Lahav:1991p833, Carroll:1992p774} apply only to the past and
fail for $a>1$. 

\subsection{Second order perturbation theory}

The perturbative solution of the equations of motion of the cosmic fluid can be written as
\begin{equation}
\delta({\bf x},a) = \sum_{J=1}^{\infty} \delta^{(J)} ({\bf x)}D^J (a)\;\;,
\label{eqn:series}
\end{equation}
where  $\delta^{(J)}D^J$ is the solution of order $J$ and we assume that the linear solution is a Gaussian random field, so all its odd-order
moments vanish. The mean value of the square of the above expansion is therefore given by a series of even powers of $D(a)$,
\begin{equation}
\langle\delta^2\rangle=   \langle[\delta^{(1)}]^2\rangle D^2   +  \langle\delta^{(2)}  + 2\delta^{(1)}\delta^{(3)}\rangle D^4  + \ldots 
\label{eqn:wick}
\end{equation}
One-loop perturbative corrections to the leading order variance $
\sigma_{L}(r) $ for power-law power spectra are given 
by \citet{lok95} and \citet{scofri96a}
\begin{equation}
\sigma^2 = \sigma^2_L + \beta \,\sigma^4_L \; ,
\label{eqn:LvsNL}
\end{equation}
where the factor $\beta$ is related to the logarithmic slope of the
two-point correlation function $ \gamma(r) = - d \ln \xi /d \ln r $ 
by the following relation:
\begin{equation}
\beta = 1.843 - 1.168 \, \gamma  . 
\label{eqn:beta}
\end{equation} 
The above equations and more generally, the weakly nonlinear gravitational instability theory were confirmed by
N-body simulations and by measurements of the galaxy skewness and
bispectrum in redshift surveys 
\citep[see e.g.][and references therein]{s3,bispec1,bispec2,verde02}. 
For non-power-law spectra, $ \gamma $ is a slowly varying function of
scale. A convenient representation 
of the correlation function over scales of interest has two power laws \citep{2002MNRAS.330..506H}, 
\begin{equation}
\xi(r) = q^2 \left( x_1^{-\gamma_1} + x_2^{-\gamma_2} \right)\, ,
\label{eqn:twoslopes}
\end{equation}
where $ x_j = r/r_j $, $ r_1 = 2.33 \Mpc $, $ r_2 = 3.51 \Mpc $, 
$ \gamma_1 = 1.72 $, $ \gamma_2 = 1.28 $, and $ q = \sigma(8\Mpc)/0.888$.
This particular choice of set of parameters are based on the PSCz
survey, and this was the choice made in the original paper,
determining \s8 from the pairwise peculiar motions of galaxies
\citep{pairwise} . We have also considered other observational estimates of $\xi(r)$ and
found that the resulting values of \s8 and $\Omega_m$ were unaffected.
The effective slope of the correlation function is then 
\begin{equation}
\gamma(r) \equiv - \frac{d \ln \xi(r)}{d \ln r} 
= \frac{\gamma_1 x_1^{-\gamma_1} + \gamma_2 x_2^{-\gamma_2}}
{x_1^{-\gamma_1} +  x_2^{-\gamma_2}} , 
\label{eqn:gamma}
\end{equation}
independent of the amplitude $q$.
At $ r = 8 \Mpc $, the effective slope is $ \gamma = 1.393 $, 
for which $ \beta = 0.216 $.
To map from $\sigma^2$ to $\sigma^2_L$, we invert  
equation~(\ref{eqn:LvsNL}): 
\begin{equation}
\sigma_L^2(r) = {\frac{\sqrt{1+4\beta\sigma^2(r)}-1}{2\beta}} \, .
\label{eqn:inversemap}
\end{equation}
Note that for $\beta  \to 0$, the above expression gives $\sigma^2_L = \sigma^2\,$, as it should. 
In \citet{pairwise} we obtained $\sigma = 1.13$. Using this value with $\beta = 0.216 $ in equation (\ref{eqn:inversemap}), we obtain for the central value 
\begin{equation}
\sigma_L = 1.02 ,
\end{equation}
a modest but significant decrease.

\subsection{Phenomenological  approach}
\label{sec:ansatz}

Our other method of relating linear and nonlinear variance is
non-perturbative. It uses the mapping of scale proposed in \citet{HKLM}, for which
conservation of mass or pair counts relates a scale in the linear regime to a nonlinear ``collapsed'' scale 
\begin{equation}
r_L^3 = \int_0^R d(r^3) \, [1 + \xi(r)] = R^3 \, [1 + \sigma^2(R)] .
\end{equation}
The variance $ \sigma^2 $ is then a (nearly) universal function 
\begin{equation}
\sigma^2(R) = g[\sigma^2_L(r_L)] \;\; .
\label{eqn:Andrew}
\end{equation}
The relation is verified and the function $g$ identified in numerical simulations \citep{HKLM,PD,Smith2003}.

There is an important difference between the perturbative and the
phenomenological calculations. The non-perturbative mapping, given by
equation~(\ref{eqn:Andrew}), explicitly uses the linear expression for
$D(a)$ and therefore in order to derive $\sigma_L$ from $\sigma$, we have
to choose specific values of the cosmological parameters; for
technical details, see, for example \citet{PD}. This is not
necessary for the perturbative formula~(\ref{eqn:inversemap}): 
to derive $\sigma_L$ we only need to know $\sigma$ and $\gamma$, the slope
of the two-point correlation function at present.

\begin{figure}
\includegraphics[width=\columnwidth]{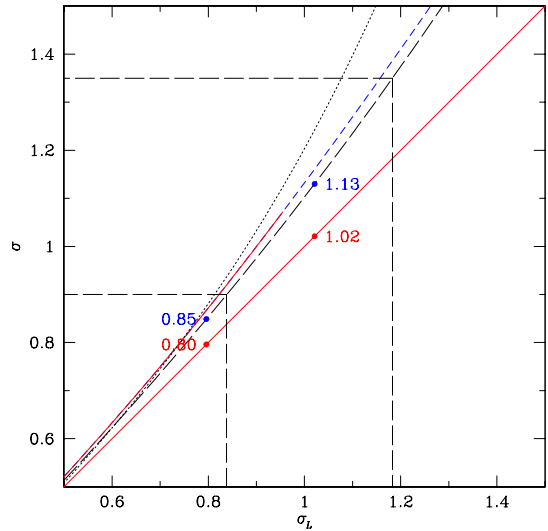}%
\caption{\label{fig:sigma8} 
Fully evolved variance $\sigma$ plotted vs. linear variance
$\sigma_L$. For comparison, 
the isolated red solid line shows $ \sigma = \sigma_L$. 
Second order perturbation theory, based on Eq.~(\ref{eqn:LvsNL}) 
with $\beta = 0.216$ gives the
long-dashed black line. The remaining curves are derived from
the phenomenological mapping, given by Eq.~(\ref{eqn:Andrew}). The differences
between them show how sensitive they are to different assumptions about the cause of
change in the amplitude of density perturbations. The growth induced
by gravitational instability acting on density perturbations with the
`standard' $\Lambda$CDM initial conditions ($\Omega_m = 0.3,
\Omega_{\Lambda} = 0.7, h = 0.7$) and the appropriate $D(a)$ factor,
is shown by the blue short-dashed line. 
The solid red line differs in
redshift or amount of evolution (and hence saturates at
$\sigma_L\approx0.95$); the dotted black line differs in Hubble
constant ($h = 0.47-1.7$) or scale ($R = 5-19$ Mpc) which 
changes $n_\eff$ and thus $\gamma$ and so tracks a little differently from the others. 
We also show the WMAP \citep{WMAP5} result ($\sigma_L =
0.80 $, which maps to $ \sigma = 0.85$), and that from measurements of cosmological flows 
($ \sigma = 1.13$ corresponding to $\sigma_L = 1.02$, with the
vertical and horizontal intervals shown in long dashed black lines the range in Eqs.~(\ref{eq:nlflow}) and (\ref{eq:lflow})).
}
\end{figure}

In Figure~\ref{fig:sigma8} we show nonlinear $\sigma$ plotted against
the inferred linear $\sigma$ (bundle of curves); 
for comparison the isolated solid red line shows $ \sigma_L $. The
perturbation theory result of equation (\ref{eqn:inversemap}) is shown
by the long-dashed black line.
To verify that perturbative results still make sense even at the
threshold of the validity of perturbation theory, when 
$ \sigma_L \approx 1 $, we also plot the phenomenological mapping
results, based on equation (\ref{eqn:Andrew}) for a variety of
parameters. In this mapping we must identify why $ \sigma_L $ has
changed, which may be from a change of fluctuation amplitude induced
by the standard gravitational instability, evolution epoch, or scale; 
each has a slightly different effect. In general, the perturbative
curve agrees well with the phenomenological results, based on
the transfer function from \cite{BBKS} and the \citet{PD} fitting. Using
the \citet{eishu98} transfer function and the \cite{Smith2003} fitting make very little difference
in the results except when the Hubble constant or scale changes. This occurs
because a change in scale substantially changes the value of $ \gamma $ and so
of $\beta$. This is yet another confirmation of the reliability of our
perturbative calculations. It also shows that the PSCz $\xi(r)$, assumed in the perturbative calculation,
agrees well with the `standard'  $\Lambda$CDM $\xi(r)$, assumed in the phenomenological mapping. 

In the phenomenological $\Lambda$CDM relation, the nonlinear signal range
\begin{equation}\label{eq:nlflow}
\sigma = 1.13^{+0.22}_{- 0.23} , 
\end{equation}
maps to 
\begin{equation}\label{eq:lflow}
\sigma_L = 1.02^{+0.16}_{-0.18} .
\end{equation}
Note that since $\sigma$ is steeper than $\sigma_L$, the range in
$\sigma_L$ is somewhat narrower 
than the one in $\sigma$. These values and the resulting limits are also shown in Fig. \ref{fig:sigma8}.

\begin{figure}[htbp]
	\centering
		\includegraphics[width=\columnwidth]{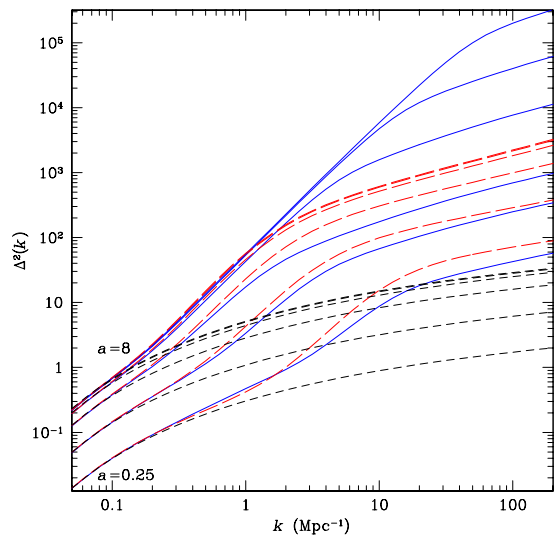}
	\caption{The linear (short-dashed black lines) and nonlinear
		values of the dimensionless $\Delta^2(k)$ as obtained from the nonlinear ansatz of \citet[long-dashed red lines]{PD}
		 and from \citet[solid blue lines]{Smith2003}, for expansion
		factors $a = 1/4$, 1/2, 1, 2, 4, and 8 (where the present is $a = 1$).}
	\label{fig:pz}
\end{figure}

\begin{figure}[htbp]
	\centering
		\includegraphics[width=\columnwidth]{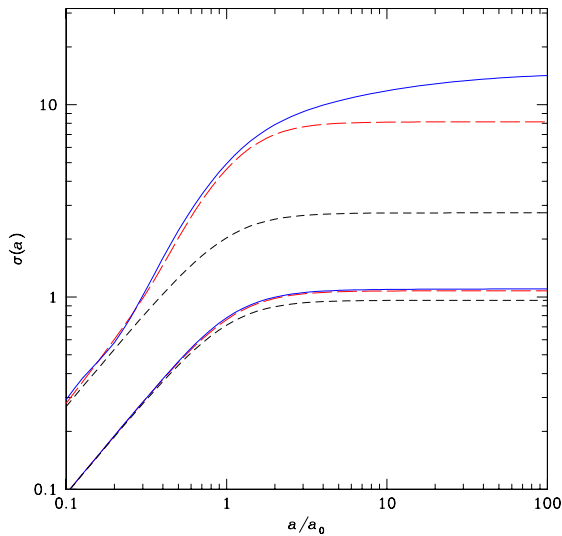}
	\caption{The value of $\sigma_8$ (lower set of curves) and $\sigma_1$ (upper set of curves)  as a function of expansion factor, for linear theory (short-dashed black lines), \citet[long-dashed red lines]{PD} and from \citet[solid blue lines]{Smith2003}.}
	\label{fig:sz}
\end{figure}

More generally, the existing descriptions and ansatzen of nonlinear
clustering were created to describe the evolution 
of clustering from the past through to the present-day; they are not
necessarily adequate representations 
of  future clustering, even in the very near future. In Figure~\ref{fig:pz} we show the value of  $\Delta^2(k,z)$,
the mean square of the dimensionless density fluctuation at various
epochs using both the \citet{PD} mapping and the \citet{Smith2003}
formula; the former becomes time independent 
whereas the actual evolution will concentrate more and more
matter in smaller haloes over time. We also see this in
Figure~\ref{fig:sz} which shows the value of both $\sigma_8$ and $\sigma_1$ as a function of expansion
factor. The latter scale of $1 h^{-1}\;\textrm{Mpc}$ is  nonlinear today and therefore already
shows a difference between linear evolution and the different
nonlinear formulae, whereas the canonical  $8 h^{-1}\;\textrm{Mpc}$ scale is just going nonlinear today.

These differences arise partly from the fact the various fitting formulae did not attempt to reproduce future clustering, but moreover, from the well-known fact that we live at a special time: in $\Lambda$CDM, linear clustering will ``saturate'' soon. Hence, any prediction such as that of \citet{PD} based on a mapping of the linear spectrum will also necessarily
saturate. Moreover, the various densities are currently evolving very rapidly. For example,  at $a = 1/2$, $\Omega_m$ was approximately what $\Omega_\Lambda$ is today; at $a = 2$, $\Omega_m$ will be approximately what $\Omega_b$ is today. This also gives us some insight into the underlying (rather than just practical) limitations of these methods. The mapping proposed by \citet{PD} essentially applies equally to all scales. However, in an evolved $\Lambda$CDM universe, large--scale dynamics are dominated by the accelerated expansion and small scales by highly nonlinear stable clustering, which is captured somewhat better in the halo model.  

\section{\label{sec:discussion}Discussion}

We will now compare different estimates of \s8.

{\bf CMB}.  The  \s8 is estimated from the CMB flutuations, observed at $z\approx1100$ and then linearly extrapolated to the present era. 
The CMB itself measures a complicated (albeit linear) functional of the power spectrum, and because of degeneracies in the determination of the CMB power spectrum from the cosmological parameters, the value of $\sigma_8$ from the CMB depends on details of the data-analysis procedure, in particular on the assumed Bayesian priors on the cosmological parameters considered. The ``recommended'' value for a flat $\Lambda$CDM model (\url{http://lambda.gsfc.nasa.gov}) is $\sigma_8=0.80\pm0.04$. However, considering different models and priors on those parameters can give variations over $0.7\lesssim\sigma_8\lesssim0.9$. 

{\bf Cosmic flows}. Observations of peculiar velocities of galaxies provide an estimate of \s8 at the present, nonlinear regime. \citet{sfi1} estimated $\sigma_8\Omega_m^{0.6}=0.52\pm0.06$ using the peculiar velocities from a sample of clusters in the SFI++ survey; \citet{ZarBraHofdaC02} using density-density and velocity-velocity comparisons found   $\sigma_8\Omega_m^{0.6}=0.45\pm0.05$ both of which translate to  $\sigma_8 \gtrsim 1$. At scales of $100\Mpc$, \citet{2008arXiv0810.3658L} found $\sigma_8=1.72\pm0.56$ from 2MASS, although they only directly measure the velocity field within $30\Mpc$, on which scales they estimate a somewhat lower fluctuation amplitude. As mentioned above, pairwise velocity analysis \citep{pairwise} estimates a present day nonlinear value of $ \sigma_8 = 1.13^{+0.22}_{-0.23} $. Recent bulk flow measurements using the best available peculiar velocity surveys \citep{WatFelHud09} require the nonlinear $\sigma_8>1.11\ (0.88)$ at a 95\% (99\%) confidence level. Thus, flow measurements give higher values of \s8 than the CMB, as expected from the ideas discussed above.
We expect more precise measurements of \s8 from cosmic flows in the near future. The new surveys are deeper, denser, have better sky coverage
\citep{sfi1,sfi2,sfierr}, and we have improved our understanding of
the distance indicators needed to extract the peculiar velocities
\citep{pairwise,radluchud04,pikhud05,sarfelwat07,watfel07,felwat08}.

{\bf Other cosmological probes}.  For comparison, we provide a table, where apart from the above estimates of \s8, we also show the values,
derived from Ly$\alpha$ observations \citep{2004ApJ...617....1T}; cluster number density measurements \citep{2008arXiv0805.2207V}; weak lensing measurements of cosmic shear \citep{2007MNRAS.381..702B}; the Sunyaev-Zeldovich effect using the ACBAR receiver \citep{ReiAdeBoc08}, and from the galaxy clustering power spectrum  \citep{2005MNRAS.362..505C,2004ApJ...606..702T,2005ApJ...633..560E}. All of these measurements fall in the range $0.7<\sigma_8<1.2$ with the most likely values in the interval $0.8<\sigma_8<0.95$. 

\begin{table*}
\caption {$\sigma_8$ from various estimators} 
\begin{center}
\begin{tabular}{lcc}
\hline \hline
Method & Parameter & value \\ \hline
CMB\footnote{}  &  $\sigma_L$& $0.80\pm0.04$\\
LY-$\alpha$ \footnote{} & $\sigma_L$  &$0.85\pm0.02$\\ 
Cosmic Shear \footnote{} &  $\sigma_L$& $0.84\pm0.05$\\
Clusters \footnote{} &  $\sigma_L$& $0.75\pm0.01$\\
SZ (ACBAR)  \footnote{} & $\sigma_8$ & $0.94^{+0.03}_{-0.04}$\\
Galaxies \footnote{} & $\sigma_8^{\rm gal}$ & $0.92\pm0.06$\\
Flows &&\\
\ Pairwise velocities  \footnote{} & $ \sigma_8 $ & $1.13^{+0.22}_{-0.23} $\\
\ Bulk flow \footnote{} &$ \sigma_8 $ &  $>1.11\ (0.88)$ at 95\% (99\%) CL\\
\hline
\label{tab:sigma8}
\end{tabular}
\end{center}
{$^1$\citet{WMAP5}; $^2$\citet{2004ApJ...617....1T}; $^3$\citet{2007MNRAS.381..702B}; $^4$\citet{2008arXiv0805.2207V}; $^5$\citet{ReiAdeBoc08}; $^6$\citet{2005MNRAS.362..505C,2004ApJ...606..702T,2005ApJ...633..560E}; $^7$\citet{pairwise}; $^8$\citet{WatFelHud09}
} 
\end{table*}

\section{\label{sec:conc}Conclusions}

We have presented a formalism to calculate  \s8, the amplitude of
cosmological density fluctuations on scales of $8\Mpc$, that 
incorporates the nonlinear evolution of the parameter. Estimates of
\s8 depend directly on the epoch and scale of the 
surveys used. When using deep, high-redshift surveys (CMB) that
estimate \s8 in the linear regime, the results suggest systematically lower
values of \s8. When analyzing shallow, local data (peculiar velocities) which are
affected by nonlinearities, we get higher \s8. The results from other
cosmological probes, shown in Table 1, show a similar trend. When results from deep surveys
are being corrected for this effect, most estimates from various independent
surveys on all scales agree better with each other: the systematic differences are reduced.

Quantitatively, our main result here is the reduction in the value of
\s8, derived from the observed mean pairwise velocity of galaxies. The
original estimate, as we have discussed earlier, is
\begin{equation}
\sigma_8 = 1.13^{+0.22}_{-0.23} \; .
  \label{eq:v12c}
\end{equation}
After the correction based on the second order perturbation theory,
this becomes
\begin{equation}\label{eq:lflowc}
\sigma_L = 1.02^{+0.16}_{-0.18} .
\end{equation}
Somewhat more cumbersome non-perturbative methods give identical
results, bringing late-time estimates of \s8 close to high-redsift
measurements, which appear in Table 1.

\noindent{\bf Acknowlegements:} 
We would like to thank the anonymous referee for her/his constructive
critical remarks. RJ was supported by the Polish Ministry of Science grant NN203 2942 34
and an INTAS grant No. 06-1000017-9258. HAF has been supported in part by
a grant from the Research Corporation, by an NSF grant AST-0807326 and
by the National Science Foundation through TeraGrid resources provided
by the NCSA. AHJ was supported by STFC in the UK.

\bibliography{sigma8}

\end{document}